\documentclass{emulateapj}

\usepackage{psfig}
\usepackage{amsmath}
\usepackage{amssymb}
\usepackage{graphicx}
\usepackage{color}
\newcommand{\lSect}[1]{{\label{sec:#1}}}

\def\gtaprx {\lower .1ex\hbox{\rlap{\raise .6ex\hbox{\hskip .3ex
{\ifmmode{\scriptscriptstyle >}\else
{$\scriptscriptstyle >$}\fi}}}
\kern -.4ex{\ifmmode{\scriptscriptstyle \sim}\else
{$\scriptscriptstyle\sim$}\fi}}}
\def\ltaprx {\lower .1ex\hbox{\rlap{\raise .6ex\hbox{\hskip .3ex
{\ifmmode{\scriptscriptstyle <}\else
{$\scriptscriptstyle <$}\fi}}}
\kern -.4ex{\ifmmode{\scriptscriptstyle \sim}\else
{$\scriptscriptstyle\sim$}\fi}}}

\newcommand{\cutt}[1]{\textcolor{blue}{}}

\newcommand{\code}[1]{{\tt{#1}}}
\newcommand{\Msun}{{\ensuremath{\mathrm{M}_{\odot} }}}
\newcommand{\Zsun}{\ensuremath{Z_\odot}}

\newcommand{\Ni}{{\ensuremath{^{56}\mathrm{Ni}}}}

\newcommand{\Co}{{\ensuremath{^{56}\mathrm{Co}}}}

\newcommand{\Sectff}[1]{{\ref{sec:#1}}}
\newcommand{\Sect}[1]{{\S~\Sectff{#1}}}

\begin{document}

\title{3D simulations of Rayleigh-Taylor mixing in
core-collapse SNe with CASTRO}

  \author{C.C.  Joggerst,\altaffilmark{1,2}
    A. Almgren,\altaffilmark{3}, S. E. Woosley\altaffilmark{1}}

\altaffiltext{1}{Department of Astronomy and Astrophysics,
University of California at Santa Cruz, Santa Cruz, CA
95060; cchurch@ucolick.org}

\altaffiltext{2}{Theoretical Astrophysics (T-2), Los Alamos National
Laboratory, Los Alamos, NM 87545}

\altaffiltext{3}{Computational Research Division, Lawrence Berkeley
National Lab, Berkeley, CA 94720}

\begin{abstract}
We present multidimensional simulations of the post-explosion
hydrodynamics in three different 15 \Msun \ supernova models with
zero, $10^{-4} \Zsun$, and \Zsun \ metallicities.  We follow the
growth of the Rayleigh-Taylor instability that mixes together the
stellar layers in the wake of the explosion.  Models are initialized
with spherically symmetric explosions and perturbations are seeded by
the grid.  Calculations are performed in two-dimensional axisymmetric
and three-dimensional Cartesian coordinates using the new Eulerian
hydrodynamics code, \code{CASTRO}.  We find as in previous work, that
Rayleigh-Taylor perturbations initially grow faster in 3D than in 2D.
As the Rayleigh-Taylor fingers interact with one another, mixing
proceeds to a greater degree in 3D than in 2D, reducing the local
Atwood number and slowing the growth rate of the instability in 3D
relative to 2D.  By the time mixing has stopped, the width of the
mixed region is similar in 2D and 3D simulations provided the
Rayleigh-Taylor fingers show significant interaction.  Our results
imply that 2D simulations of light curves and nucleosynthesis in
supernovae (SNe) that die as red giants may capture the features of an initially spherically symmetric explosion in far less computational time than required
by a full 3D simulation.  However, capturing large departures from
spherical symmetry requires a significantly perturbed explosion.
Large scale asymmetries cannot develop through an inverse casacde of
merging Rayleigh-Taylor structures; they must arise from asymmetries
in the initial explosion.

\end{abstract}


\maketitle

\section{Introduction}
\lSect{introduction}

Supernova 1987A has furnished astronomers with perhaps the best
opportunity to study a core-collapse explosion to date.  One of the
most exciting discoveries to emerge from this event is the evidence
for large-scale, extensive mixing in the supernova ejecta.
$\gamma-$ray lines emitted by the decay of \Co \ were detected half a
year earlier than expected for a spherically symmetric explosion model
\citep{Matz1988}.  Modeling the light curve in 1D requires a large
amount of \Ni \ to be mixed outward and H and He to be mixed inward
\citep{Utrobin2004}.  The Bochum event, in which fine-structure
H$\alpha$ lines are observed 2 weeks after the explosion, is explained
by the ejection of $10^{-3}$ \Msun \ of \Ni \ moving with a velocity
in excess of 4000 km s$^{-1}$ into the far hemisphere of 1987A
\citep{Hanuschik1988}.  The Rayleigh-Taylor (RT) instability was
posited as a possible explanation for the large-scale mixing implied
for this explosion.  Many groups using both SPH and grid-based codes
simulated the evolution of 87A-like progenitor models, first in 2D,
then in 3D.  It soon became apparent that more than just the Rayleigh-Taylor
instability is needed to explain the high \Ni \ velocities as well as the
large degree of outward mixing of heavy elements and inward mixing of
lighter elements.  Simulations that posit modest initial perturbations
are unable to replicate these features of the explosions.
\citet{Kifonidis2006} find that by following the explosion from early
times, and using large, low-order perturbations in the inner layers,
they are able to replicate the high \Ni \ velocities and large amounts
of mixing seen in the explosion.  The standing accretion shock
instability (SASI) \citep{Blondin2003, Blondin2006, Scheck2006,
  Burrows2006, Burrows2007, Marek2009} provides a mechanism by which
the inner parts of the ejecta could be deformed.

Most studies of the Rayleigh-Taylor instability in core collapse supernovae use
1987A-like progenitor models.  However, supernovae in the high-redshift
universe, as well as many modern day supernovae which die as red, not
blue, giants, explode with presupernova structures which differ
significantly from 1987A. This may greatly influence the evolution of
the Rayleigh-Taylor instability in their outer layers.  Simulations of rotating
zero-metallicity supernovae \citep{Ekstroem2008, Hirschi2007} indicate
that they may die as extended red, rather than compact blue, giants.
Rotation mixes CNO elements to the base of the hydrogen-burning shell,
where they catalyze the CNO cycle, greatly boosting the burning rate
of H.  This leads to the formation of a large convective envelope,
effectively turning a compact blue star into a giant red one.  Models
indicate that the shell boost induced convection effectively mixes
together the helium shell and the hydrogen envelope.  The dense helium
shell has been found to be important to the post-bounce evolution of
1987A-like progenitors \citep{Hammer2010}.  Rotation has less of an
effect on early stars with very low, but not zero, metallicity, such
as the Z = $10^{-4}$ \Zsun \ progenitor models studied in
\citet{Joggerst2010} as well as the current work. These stars remain
blue.  Solar metallicity stars end their lives as red giants, but
retain their helium shells. Red stars should evolve differently from
1987A-type models after the supernova shock is launched.  They are
larger, so the shock takes longer to traverse the star, and the Rayleigh-Taylor
instabilities have more time to develop \citep{Chevalier1976}.

Previous simulations of core-collapse explosions in two dimensions
\citep{Arnett1989, Hachisu1990, Fryxell1991, Herant&Benz1991,
  Mueller1991, Herant&Benz1992, Hachisu1992} have primarily focused on
87A-like progenitor models. Studies of Rayleigh-Taylor-induced mixing in three
dimensions has exclusively focused on attempts to replicate
observations of 1987A, and have found significant differences between
simulations performed in two and three dimensions.  The most
successful of these 2D attempts, \citet{Kifonidis2003, Kifonidis2006}
found that to reproduce the high \Ni \ velocities observed in 1987A
large initial asymmetries of low mode order were necessary. In studies
of mixing in both red and blue supergiants\citep{Joggerst2009,
  Herant&Woosley} mixing is found to be more vigorous in red stars
than in blue stars.  \citet{Joggerst2010} evolve 36 models spanning
three rotation rates, three explosion energies, three masses, and two
metallicities in two dimensions, finding that mixing is more vigorous
and goes on longer in stars that die as red as opposed to blue giants,
in higher energy explosions, and in less massive stars.  The
literature on 3D simulations is more sparse.  \citet{Kane2000} find
that for a single-mode perturbation of $10\%$ amplitude, Rayleigh-Taylor fingers
grow $30-40\%$ faster in 3D than in 2D.  The most recent paper on the
post-explosion hydrodynamics in 1987A \citep{Hammer2010} use a
successful explosion model from \citet{Scheck2007} to initialize a 3D
simulation, along with several 2D simulations derived from meridional
slices through the initial explosion model.  They find that for this
model, which is highly asymmetric in the inner regions, the Rayleigh-Taylor
instability grows faster and ballistically moving clumps of \Ni
\ reach the hydrogen envelope in 3D. These clumps are effectively
stopped by the dense He shell in the suite of 2D models initialized from
slices through the \citet{Scheck2007} model.

This is the first paper to simulate the post-explosion hydrodynamics
of models with a diversity of presupernova structures in three
dimensions. All models have the same mass (15 \Msun) and explosion
energy ($1.2 x 10^{51}$ ergs at infinity), but two models explode as
red giants and one as a more compact blue star.  The red giant with
zero metallicity effectively lacks a He shell because of convection
arising from the hydrogen shell boost.  The Z = $10^{-4}$ \Zsun \ and
Z= \Zsun \ models both have helium shells, though the former is blue
and the latter red at the time of explosion.  This diversity of models
allows us to see if presupernova structure has an impact on the
subsequent evolution of nearly isotropic 2D as opposed to 3D
models. These calculations set the stage for further study of
asymmetric explosions in these diverse presupernova models.

The numerical methods employed in this study are discussed in Section
\Sect{codes}, and initial models and problem configuration are discussed in Section
\Sect{setup}.  Results are presented in Section \Sect{results}, and
are discussed and compared with previous 3D simulations in
\Sect{discussion}.  We offer some conclusions in
\Sect{conclusions}.

\section{Numerical Algorithms}
\lSect{codes}

The multidimensional simulations described in this paper are performed
with \code{CASTRO} \citep{Almgren2010}, an Eulerian adaptive mesh
refinement (AMR) hydrodynamics code.  Time integration of the
hydrodynamics equations in \code{CASTRO} is based on an unsplit
higher-order piecewise parabolic method (PPM); the additional features
are described below.  All simulations described here used either
axisymmetric coordinates in 2D or Cartesian coordinates in 3D.

\subsection{Equation of State}

We followed sixteen elements, from hydrogen through {\Ni}, so that our
elemental abundances could later be used to compute spectra.  The
atomic weights and amounts of the elements are used to calculate the
mean molecular weight of the gas required by the equation of state.
The abundance of {\Ni} was followed separately in order to calculate
the energy deposited by radioactive decay of {\Ni} and \Co.

The equation of state in our simulations is as described in
\citet{Joggerst2010}. It assumed complete ionization and included
contributions from both radiation and ideal gas pressure:
\begin{equation}
        P= f(\rho,T) \, \frac{1}{3} a T^4 + \frac{k_{B} T \rho}{m_p \mu}
\end{equation}
\begin{equation}
       E = f(\rho, T)\, \frac{a T^4}{\rho} + 1.5\frac{k_{B} T}{m_p \mu},
\end{equation}
where $P$ is the pressure, $a$ is the radiation constant, $k_{B}$ is
Boltzmann's constant, $T$ is the temperature, $\rho$ is the density,
$m_p$ is the proton mass, $\mu$ is the mean molecular weight, and $E$
is the energy.

The function $f(\rho, T)$ is a measure of the contribution of
radiation pressure to the equation of state that accounts for the fact
that radiation ceases to be trapped at some point.  It is 1 in regions
where radiation pressure is important, i.~e. where gas is optically
thick, and 0 in regions where radiation pressure is unimportant,
i.~e. where gas is optically thin, with a smooth transition in
between.  The function $f(\rho,T)$ takes the form

\[ f(\rho,T) = \left\{ \begin{array}{ll}
                           0 & \mbox{if $\rho \ge 10^{-9}$ gm
cm$^{3}$} \\ & \mbox{or $T \le T_{neg}$} \\
f(T)=e^{\frac{T_{neg}-T}{T_{neg}}} & \mbox{if $\rho < 10^{-9}$ gm
cm$^{3}$}\\ & \mbox{and $T > T_{neg}$}\\
                          \end{array}
                  \right.\vspace{0.1in} \]

where $T_{neg}$ is the temperature at which contributions to the
pressure from radiation are 100 times less than that contributed by
ideal gas pressure:
\begin{equation}
  T_{neg}={\frac{3 k_b \rho }{100 m_p \mu a}}^{1/3}.
\end{equation}

Radiation pressure will begin to dominate the equation of state above
$T_{neg}$ without some adjustment, even though this is unphysical, as
radiation pressure is negligible in optically thin regions.  Damping
the radiation component of the EOS with the function $f(\rho,T)$ in
these regions provides a more physical solution.

\subsection{Radioactive Decay of \Ni}
Energy from the radioactive decay of $^{56}$Ni to $^{56}$Fe was
deposited locally at each mesh point, in the same manner as described
in \citet{Joggerst2010}.  The rate at which energy from the decay of
\Ni \ to \Co \ was deposited in the grid is given by:
\begin{equation}
               dE_{Ni} = \lambda_{Ni} X_{Ni}
               e^{-\lambda_{Ni}t}q(Ni).
\end{equation}
The decay rate of \Ni, $\lambda_{Ni}$, is $1.315\times10^{-6}$
s$^{-1}$, and the amount of energy released per gram of decaying \Ni
\ is $q(Ni)$, which we set to $2.96 \times 10^{16}$ erg g$^{-1}$. 
$X_{Ni}$ is the mass fraction of \Ni \ in the cell.  The mass
fraction of \Co \ at a given time can be expressed in terms of
the mass fraction of initial \Ni \ by
\begin{equation}
               X_{Co} =
               \frac{\lambda_{Ni}}{\lambda_{Co}-\lambda_{Ni}}
               X_{Ni} (e^{-\lambda_{Ni}t}-e^{-\lambda_{Co}t}),
\end{equation}
so that the energy deposition rate from \Co \ is
\begin{equation}
               dE_{\o} =
               \frac{\lambda_{Ni}}{\lambda_{Co}-\lambda_{Ni}}
               X_{Ni} (e^{-\lambda_{Ni} t}-e^{-\lambda_{Co} t}))
               \lambda_{Co} q(Co).
\end{equation}
We assumed a decay rate $\lambda_{Co} = 1.042 \times10^{-7}$
s$^{-1}$ and an energy per gram of decaying \Co, $q(Co)$, equal
to $6.4 \times 10^{16}$ erg g$^{-1}$.

\subsection{Gravity}

Although \code{CASTRO} supports several different approaches to
solving for self-gravity, we used the monopole approximation for
gravity in the calculations presented here, as discussed in
\citet{Joggerst2010}.  Because the density profiles in the simulations
depart very little from spherical symmetry, the gravitational field
constructed using the monopole approximation is nearly identical to
that found by solving the full Poisson equation for the gravitational
potential, and using the monopole approximation significantly reduces
the computational cost of the calculations.

Gravity from a point mass located at the origin was also included in
the gravitational potential.  The point mass represents the compact
remnant left behind by the SN explosion.  As infalling matter crosses
the zero-gradient inner boundary near the origin, it is added to this
point mass.

\subsection{Adaptive Mesh Refinement}
\lSect{AMR}

\code{CASTRO}'s AMR algorithm uses a nested hierarchy of
logically-rectangular grids with simultaneous refinement of the grids
in both space and time.  The integration algorithm proceeds
recursively, advancing coarse grids in time, then advancing fine grids
multiple steps to reach the same time as the coarse grid, and finally
synchronizing the data at different levels.

The regions of refinement evolve throughout the simulation based on
user-specified refinement criteria applied to the solution.  In the
simulations presented here, regridding of all grids at level $\ell+1$
and above occurred every two level $\ell$ time steps.

Our refinement criteria are based on the ``error estimator'' of
\citet{Lohner1987}, which is essentially the ratio of the second
derivative to the first derivative at the point at which the error is
evaluated.  Details of the implementation are discussed more fully in
\citet{Almgren2010}. The result is a dimensionless, bounded estimator,
which allows arbitrary variables to be subjected to the same criterion
for error estimation.  In these calculations, density, pressure,
velocity, and the abundances of \Ni, He, and O were used as refinement
variables, with the elemental abundances only used in a particular
region if their abundance was greater than $10^{-3}.$

\section{Simulations Setup and Execution}
\lSect{setup} The simulations presented here were carried out in two
distinct stages.  First each stellar model was evolved in one
dimension using the code \code{KEPLER} to the point where the core
became unstable to collapse.  It was then artificially exploded by
means of a piston located at the base of the oxygen shell with enough
energy that the explosions had $1.2 \times 10^{51}$ ergs of energy at
infinity.  The models were evolved forward in one dimension for 20
seconds for models z15 and u15, and 100 seconds for s15, until all
nuclear burning had ceased.

These profiles were then mapped onto a two-dimensional r-z or
three-dimensional Cartesian grid in \code{CASTRO}, and the calculation
was evolved until the shock reached the edge of the simulation domain.
The calculation was then stopped, the domain enlarged, and the
calculation restarted within the larger domain (see \Sect{embiggening}
below).  Data to fill the new regions of the enlarged domain were
supplied from the same one-dimensional profile that was used to
initialize the multidimensional calculation.

This process continued until the models had evolved out to radii where
Rayleigh-Taylor mixing ceased and the star was expanding essentially
homologously.  We computed three 15 \Msun \ models, at zero, $10^{-4}$\
\Zsun, and solar metallicities.

\subsection{Progenitor Models}
\lSect{models}
\begin{figure*}
\centering
\plotone{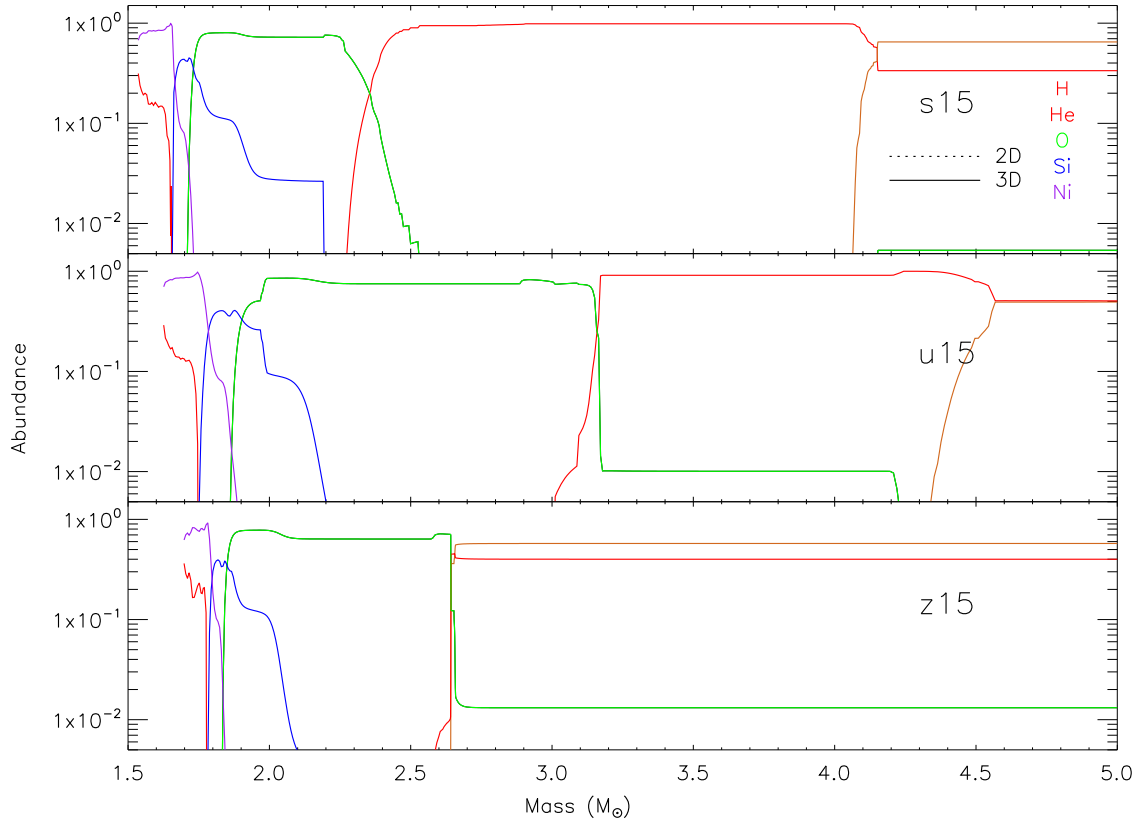}
\caption{Shell structure for H, He, O, Si, and Ni for our initial
models.  Other elements were included in our simulations but are left
out of the figure for clarity.  Note that while s15 and u15 (top and
middle figures) show an He shell, this shell is not present in model
z15 because of convection arising from a hydrogen shell boost.  The
models are 15 solar masses, but the outer layers are essentially a
continuation of the levels at the right hand side of the plot, and
have been omitted to better show the detail in the center.}
\label{fig:initial}
\end{figure*}

Three different supernova models, representing three different
metallicities, are presented in this paper.  Model z15 has zero
initial metallicity, and represents a Population III (Pop III) star.
Model u15 has Z = $10^{-4}$ \Zsun, and model s15 has solar
metallicity. All models are derived from rotating 15 \Msun \
progenitors.  The zero and low metallicity models presented in this
paper are the same as presented in \citet{Joggerst2010} and
\citet{Zhang2008}.  The solar metallicity model, model s15, is similar
to the model presented in \citet{Woosley&Heger2007}.  For moderate
values of rotation, the amount of rotation was found to have little
effect on its post-explosion evolution \citep{Joggerst2010}.  There
are, however, profound differences between rotating and non-rotating
zero metallicity supernovae.  Several studies \citep{Ekstroem2008,
Hirschi2007, Joggerst2010}) have found that a small amount of rotation
in zero metallicity massive stars dredges up enough carbon, nitrogen,
and oxygen from the helium burning shell to catalyze a CNO cycle at
the base of the hydrogen envelope.  This leads to rapid burning and a
shell boost that triggers convection, mixing most of the helium and
hydrogen shells together, and ``puffing up'' the outer envelope of the
star.  This is apparent in Figure \ref{fig:initial}.  Models with a
small amount of metals (i. e. u15 in this paper) do not show this
shell boost behavior, and die as blue supergiants.  Rotation also has
little effect on the presupernova structure of the solar metallicity
model s15, which dies as a red giant.

\subsection{Initialization of Multidimensional Data}

In mapping the radial data from \code{KEPLER} onto the
multidimensional grids in \code{CASTRO}, special care was taken to
properly resolve the key elements of the simulations: the shock, the
elemental shells, and the \Ni \ at the center of the explosion. In
particular, both the \Ni \ and the O shell were resolved with a
minimum of 16 cells at the highest level of refinement.  This mapping
results in explosions that are spherically symmetric in \code{CASTRO};
low-order departures from spherical symmetry are suppressed. Our
models therefore only capture higher-order asymmetries in the
explosions.

Perturbations that give rise to the Rayleigh-Taylor instabilities are seeded from
the axisymmetric or Cartesian grid. We performed calculations in 2D at
twice the resolution of the simulations considered in this paper and
compared the results of the high resolution simulations to the low
resolution simulations.  The width of the mixed region and velocities
of the elements were essentially the same, indicating that our results
are numerically converged and do not depend on the initial scale of
perturbations imposed by the grid.

\subsection{Enlarging the Domain}
\lSect{embiggening}

In order to minimize the amount of computational resources used, we
implemented a strategy of enlarging the simulation domain only as
necessary as the size of the region of interest increased in time.
The simulations were initialized on a $128^{n}$ grid with 2 levels of
refinement, where $n$ is the dimensionality of the simulation.  This
gives an effective grid resolution of $512^{n}$ at the finest level.
A given simulation was advanced in time to the point where the shock was
near the edge of the domain.  The simulation was then stopped and the
domain was doubled in each coordinate dimension.  The enlarged domain
was covered by grids with cell spacing twice that of the previous
coarsest resolution, so that after each enlargement the base grid
continued to be $128^n.$ The simulation then had 3 levels of
refinement; in most cases (except for u15) the highest level of
refinement was removed at this point, reducing the finest resolution
by a factor of 2 as well. This allowed us to tailor the size of the
domain to the size of the region of interest, without wasting
computational resources in regions where nothing of interest was
happening.  Each time the domain was enlarged data from the new
regions of the domain were interpolated from the original model.  
 
To ensure that the strategy described above did not affect the results
of the simulations relative to analogous calculations performed on
domains which were larger throughout, we ran two-dimensional
simulations in which the final domain size was imposed from the start.
The results were essentially the same, although more complex mixing
structures are visible in the case with the initially larger domain.
The width of the mixed region remains the same, i.e. heavy elements
are mixed out the the same point in radius and mass, while lighter
elements are mixed inward to the same point in the both cases.  The
velocities obtained by the elements are the same.  This is sufficient
evidence that the strategy of enlarging the domain throughout the
calculation did not introduce spurious artifacts.

\section{Results}
\lSect{results}

In each multidimensional simulation presented here, when the shock
encountered the large region of increasing $\rho r^{3}$ at the base of
the hydrogen envelope of the presupernova star (or, in the case of
model z15, the helium-hydrogen envelope) it slowed, giving rise to a
reverse shock.  The reverse shock then propagated inwards in mass
towards the center of the star, and left a reversed pressure gradient
in its wake.  This triggered the Rayleigh-Taylor instability first
between the helium shell and the hydrogen envelope, and then between
the helium and oxygen layers of the supernova.  For model z15, this
instability first arose between the oxygen layer and the
helium-hydrogen envelope.  The instability grew, mixing together
layers of increasing atomic mass as the reverse shock propagated inward
through the star. Eventually the reverse shock passed by, and mixing
stopped once the Rayleigh-Taylor fingers had dissipated their momenta.

In each simulation, the domain was enlarged five times, bringing the
final effective resolution of each simulation to $16,384^{n}$,
where $n$ is the dimensionality of the simulation.  For model u15,
four levels of refinement were retained at the final stage of the
simulation; for models s15 and z15, two levels of refinement were
retained.

Mixing had effectively ceased by $4.0x10^{5}$ seconds for model s15,
$3.8x10^{4}$ seconds for model u15, and $1.7x10^{5}$ seconds for model
z15.  Mixing had stopped by similar times in 2D and 3D simulations.
The models were run to twice this time to make sure that no additional
mixing would occur.

\begin{figure*}
\centering
\plotone{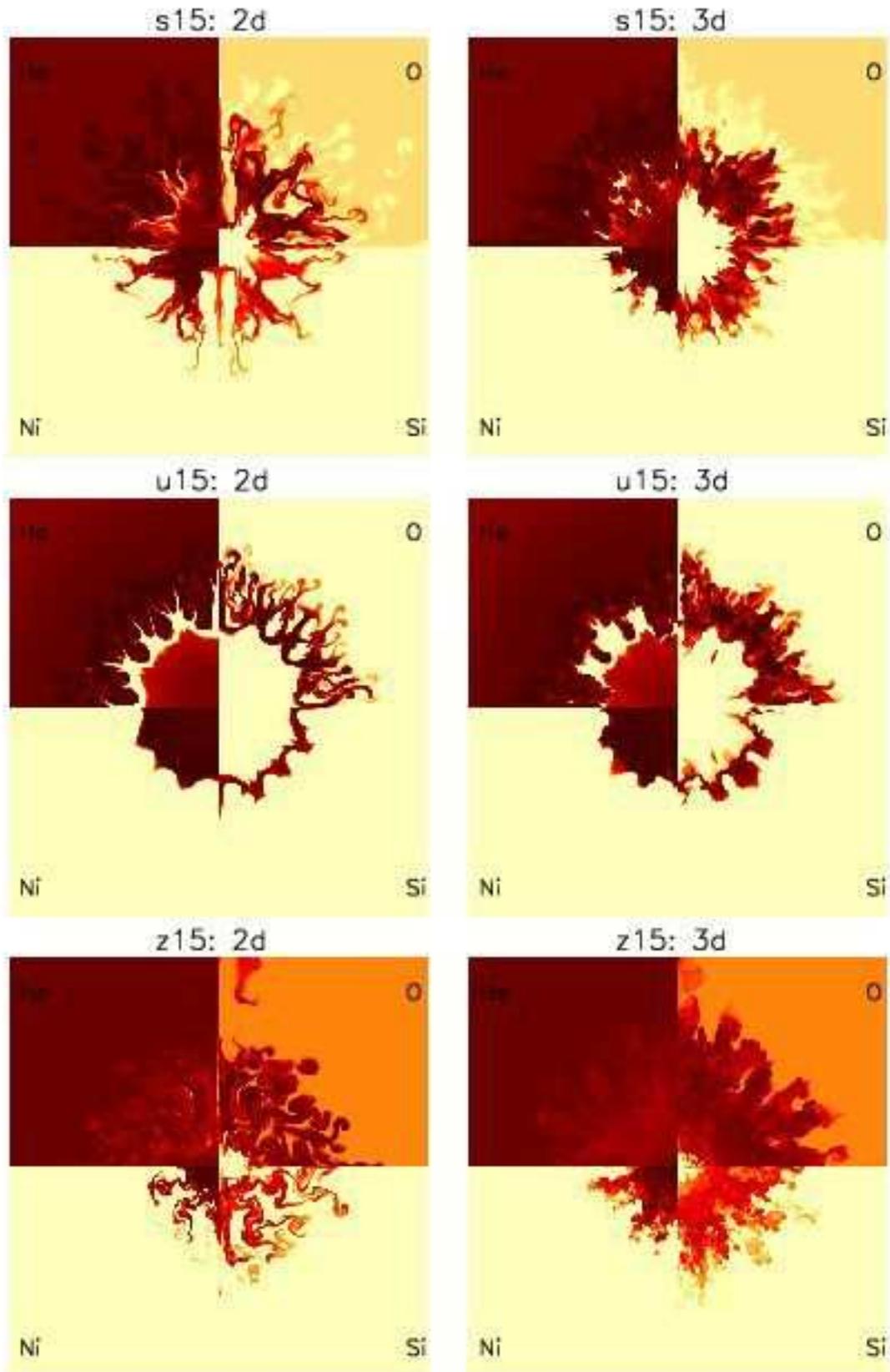}
\caption{Snapshots of elemental abundance in the 2D (left) and 3D
  (right) versions of (from top to bottom) models s15, u15, and z15.
  The 3D snapshots are slices at the X axis through the YZ plane.
  Because of artifacts in CASTRO, the extent of mixing is likely to be
  slightly exaggerated in slices at an axis.  Nevertheless, the width
  of the mixed region in 2D and 3D appear very similar in all models.
  There is even slightly less mixing of \Ni \ in model s15 in 3D than
  in 2D.  The shape of the instability differs slightly between 2d and
  3d in the expected manner--the 2D shapes are slightly more
  ``mushroomed'' and the 3D shapes are more elongated, with the 3D
  models appearing to have transitioned to a fully turbulent regime,
  but otherwise the two simulations appear very similar.}
\label{fig:2d3dsnap}
\end{figure*}

Shown in Figure \ref{fig:2d3dsnap} are snapshots at identical times
for the two- and three-dimensional simulations of models (from top)
s15, u15, and z15.  The abundance of He, O, Si, and \Ni \ are shown in
a logarithmic scale, starting clockwise from top left.  These
snapshots correspond to times when mixing had effectively stopped in
the different models.  On the left are snapshots of the 2D models; at
right are snapshots showing slices in the YZ plane at the X axis.
Upon first inspection, there appears to be very little difference
between the two and three dimensional simulations of all three models.
The width of the mixed region appears the same between the 2D and 3D
models, although mixing in the 3D models appears to be more complete
in this region than in the 2D models.  The shape of the instability
differs slightly between the 2D and 3D simulations. The mushroom shape
of the instability is more clearly defined in 2D than in 3D,
especially for model u15, where the instability had less time to grow
and thus retained its original shape more clearly.  In the 3D models,
the Rayleigh-Taylor instability appears more elongated, in line with what was found
in the single mode study of \citet{Kane2000}.  The greatest difference
between the 2D and 3D simulations arises in model s15, shown at the
bottom of Figure \ref{fig:2d3dsnap}, where there is actually
\emph{less} mixing of \Ni \ in 2D than in 3D.

In all models the Rayleigh-Taylor fingers have grown to the point where they have
begun to interact with one another.  Models z15 and s15, in which the
RT instability had the longest time to grow, show the greatest degree
of interaction.  This interaction transfers energy and momentum in the
transverse directions to the blast wave, leading to a transition to
turbulence in 3D, and to a chaotic regime in 2D \citep{Remington2006}.
\citet{Miles2005} state that this transition to turbulence begins to
happen when the Rayleigh-Taylor fingers are about 5 to 6 times longer than their
initial wavelength.  This has clearly happened in all simulations.
Models z15 and s15 appear strikingly similar to the fully turbulent
simulations presented in \citet{Miles2005}.

\begin{figure*}
\centering
\plotone{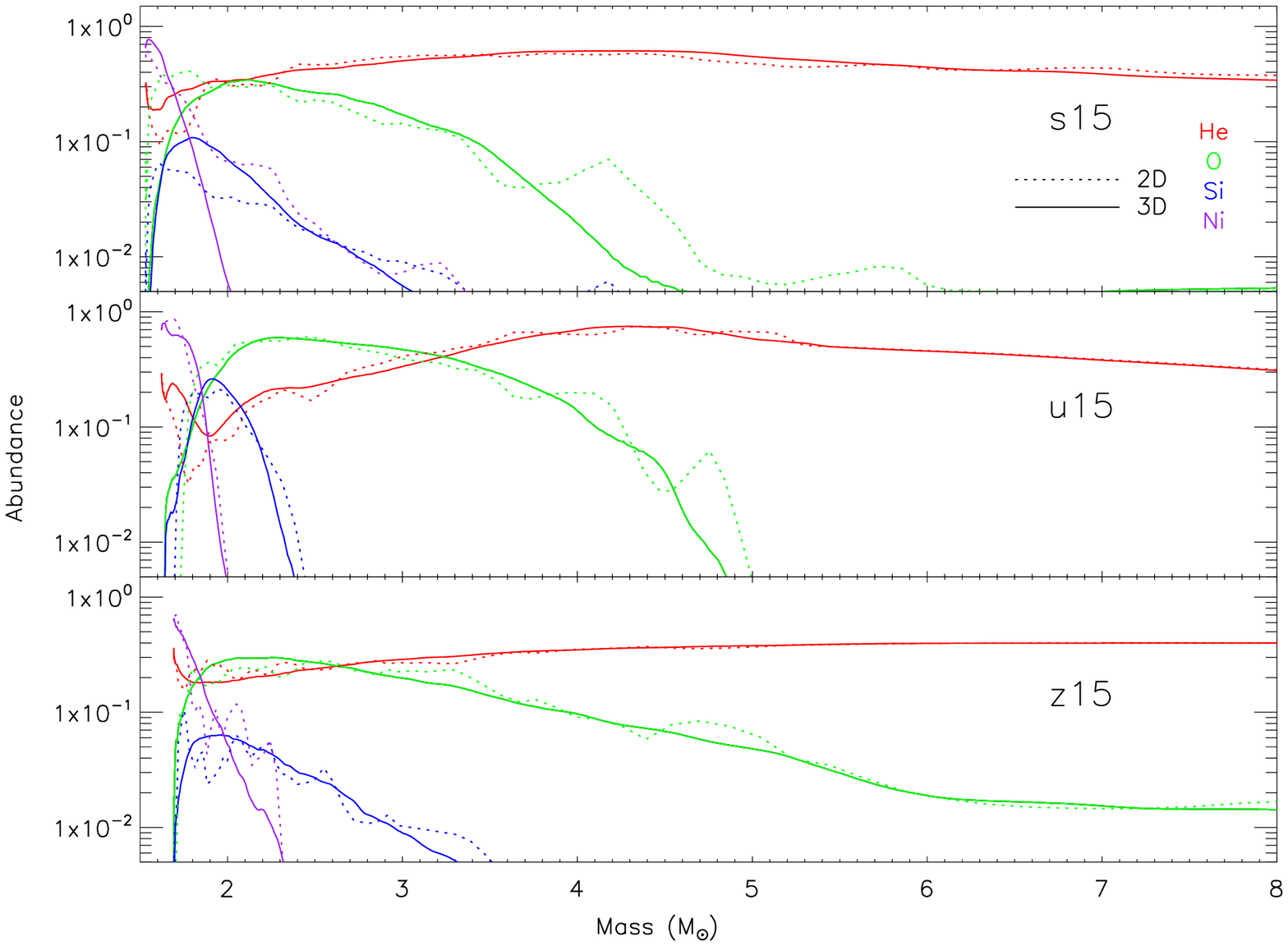}
\caption{Abundance by mass of individual elements as a function of
  mass coordinate.  The width of the mixed region is essentially the
  same between the 2D and 3D cases, though 3D is smoother. The higher
  mass coordinate layers of the simulations have been omitted from the
  plot to better show the details of their inner portions.}

\label{fig:mpmp}
\end{figure*}

The differences between 2D and 3D can be compared more quantitatively
by examining Figure \ref{fig:mpmp}.  This figure displays the radial
average of abundances for significant elements in both 2D and 3D
simulations of the three models after all mixing has stopped.  The
times shown are the same times as in Figure \ref{fig:2d3dsnap}. Figure
\ref{fig:mpmp} shows clearly that the width of the mixing region in
the 2D and 3D models is nearly the same.  \Ni \ is the exception, and
is slightly more mixed in 2D than in 3D in the simulations of model
s15.  Previous simulations of the Rayleigh-Taylor instability in which these
fingers do not interact show $\approx$ 30$\%$ faster growth rate in 3D
than in 2D, resulting in a wider mixed region in 3D.  We also see a
faster initial growth rate in 3D than in 2D, but once the Rayleigh-Taylor fingers
begin to interact with one another, the growth rate in the 3D models
decreases and the final width of the mixed region is the same between
2D and 3D.  This is in line with simulations of the Rayleigh-Taylor instability
in a laboratory context presented in \citet{Miles2005}, in which the
RT fingers interacted.
 
The 2D and 3D simulations do appear somewhat different in Figure
\ref{fig:mpmp} in that the 3D lines are smoother than the 2D lines.
Because interactions between Rayleigh-Taylor fingers can transfer momentum and
energy perpendicular to the direction of the blast wave, mixing
proceeds to a greater extent in 3D than in 2D.  This reflects the fact
that the 3D simulations have become fully turbulent (in the case of
z15 and s15), while the 2D simulations are chaotic, as 3D simulations
are able to transfer momentum and energy in two dimensions transverse
to the blast wave, while 2D simulations have only one transverse
dimension at their disposal.  There is likely a sampling effect at
play here, as well.  In 2D, there simply are not as many Rayleigh-Taylor fingers as
there are in 3D, and thus there is effectively a smaller sample of the
instabilities that grow.  In 3D there are more Rayleigh-Taylor structures, so the
space the fingers can occupy is better sampled, leading to a smoother
radialized distribution.

\begin{figure*}
\centering
\plotone{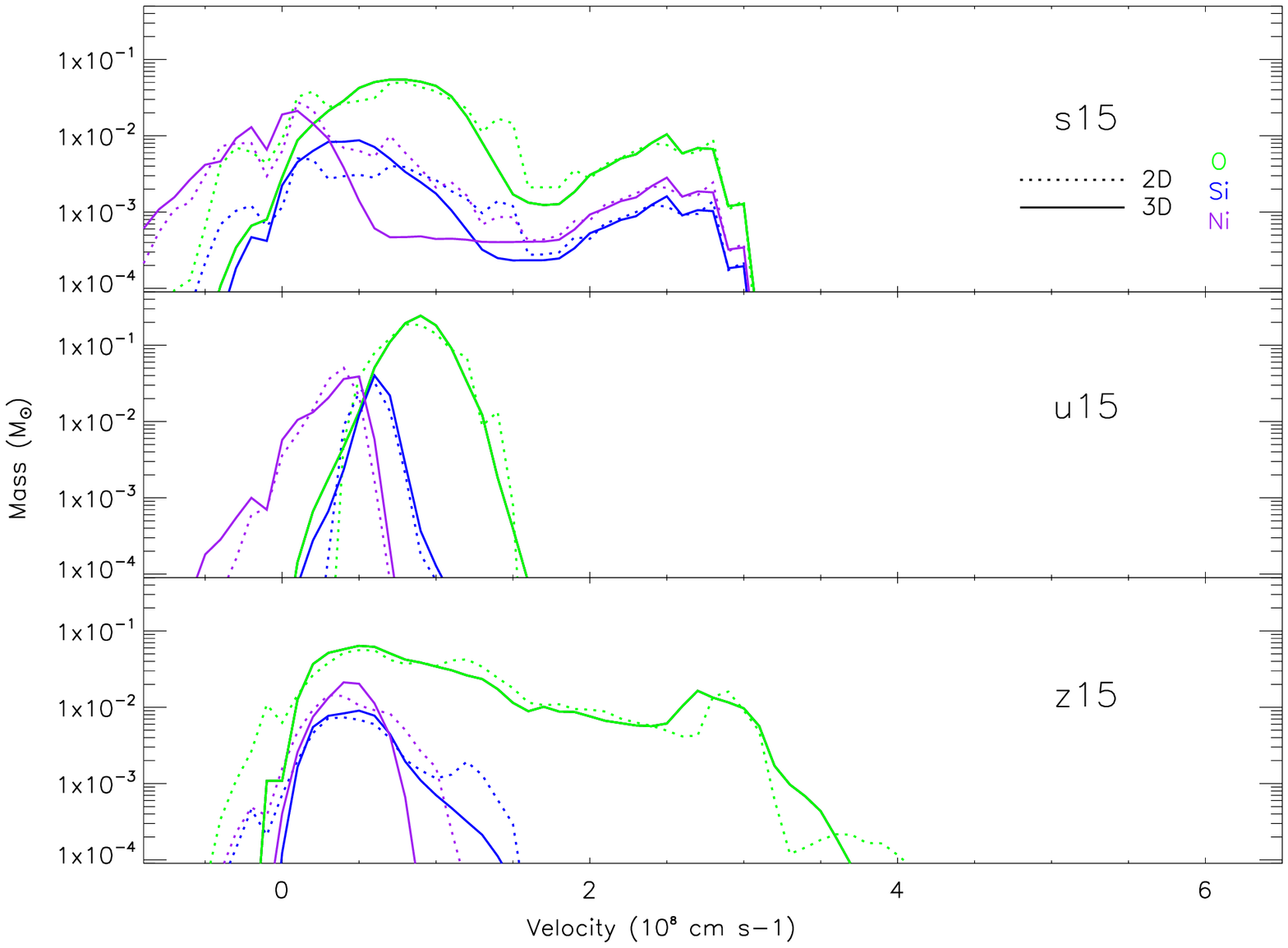}
\caption{Mass of individual elements as a function of velocity per 100
  km s$^{-1}$ velocity bin.  The amount of heavier elements such as O
  and \Ni \ at high velocities is essentially the same in 2D and 3D.
  Where the two differ, as they do most significantly for Ni and Si in
  models s15 and z15, the 2D simulations actually show material mixed
  out to higher velocities than the 3D simulations.}
\label{fig:mpvp}
\end{figure*}

The velocities obtained when mixing has effectively ceased are
virtually identical between the 2D and 3D simulations, as shown in
Figure \ref{fig:mpvp}. These simulations are shown at different times:
$4.0x10^{5}$ seconds, $3.8x10^{4}$ seconds, and $1.7x10^{5}$ seconds
for models s15, u15, and z15, respectively, and so direct comparison
of one model with another would be misleading, as the material will
slow down as time goes on.  It is clear from Figure \ref{fig:mpvp}
that the ultimate velocities after mixing has stopped and the Rayleigh-Taylor
instability has frozen out are essentially the same in the 2D and 3D
simulations.  This is different from what \citet{Hammer2010} found in
their simulations with larger degrees of initial asymmetry.  They saw
a high velocity tail for \Ni, O, and Si.  We do not see a high
velocity tail here.

\begin{figure*}
\centering
\plotone{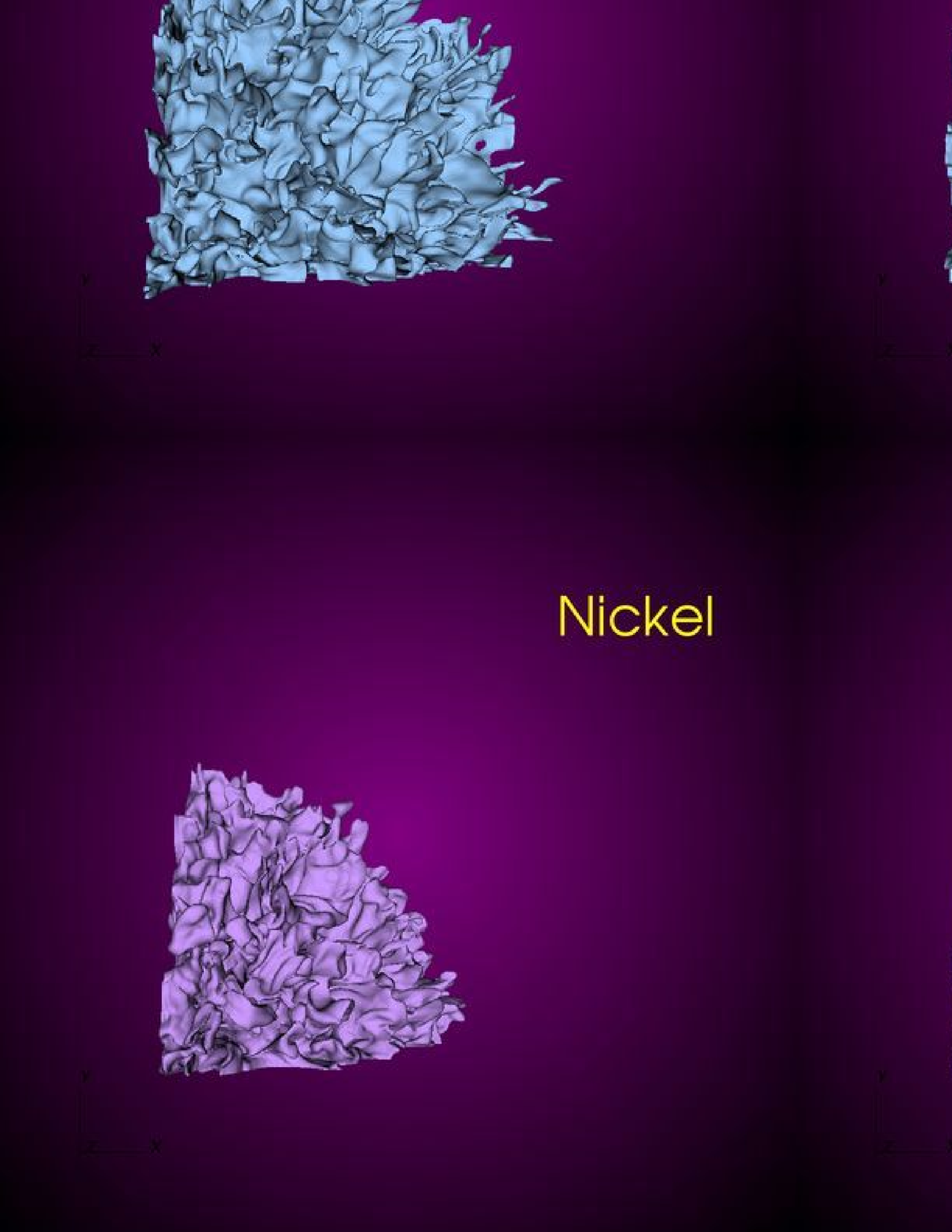}
\caption{3D renderings of the outermost surface of carbon, oxygen,
silicon, and nickel after mixing has frozen out. At left is model s15;
center is u15, and right is z15.  Note that while the scale and time
are consistent within models, the 3 models are shown at different
scales and time: $5 x 10^{13}$ cm and $4.0x10^{5}$ seconds for s15, $5x
10^{12}$ cm and $3.8x10^{4}$ seconds for model u15, and $3 x 10^{13}$
cm and $1.7x10^{5}$ seconds for model z15. For no model did heavier,
inner layers penetrate the lighter, outer layers, though for model z15
clumps of material have broken off and moved a small distance into the
H-He envelope. }
\label{fig:rend}
\end{figure*}

The 3D renderings shown in Figure \ref{fig:rend} give a clearer
picture of the shape of the Rayleigh-Taylor instabilities in our simulations than
the slice through the 3D plane shown in Figure \ref{fig:2d3dsnap}.  We
show renderings of the outermost surfaces of carbon, oxygen, silicon,
and nickel. Model s15 is shown at the left, model u15 in the center,
and model z15 at the right.  These renderings were taken at different
times, but after mixing had frozen out.  Only in model z15 have bits
of the oxygen shell broken off and begun to move through the H-He
shell that surrounds the heavy element core of this star.  For no
models have clumps of heavier elements penetrated shells of lighter
elements, as happened in the asymmetric explosion modeled by
\citet{Hammer2010}.  Even the broken-off clumps of oxygen visible in
the right hand panel of Figure \ref{fig:rend} are encased in a layer
of carbon.  It is also apparent by examining this figure that mixing
has progressed further in the simulations where the progenitor star
was a red giant, model s15 and z15, than for model u15, whose
progenitor died as a more compact blue star.  The original shape of
the instability can be seen in the carbon surface (top middle panel).
In addition, the \Ni \ at the center of the model is less deformed in
this model than in previous models.  The most vigourous mixing has
taken place in model z15, which lacked a dense helium shell and died
as a red giant.

\section{Discussion}
\lSect{discussion} 

%discussion of previous simulations in the same vein
Previous simulations of the Rayleigh-Taylor instability in core
collapse supernovae have found significant differences between similar
calculations performed in two and three dimensions.  \citet{Kane2000}
examined the growth of Rayleigh-Taylor fingers arising from a single mode
perturbation, and found that the instability grew $30-35\%$ faster in
3D than in 2D in a 1987A-type progenitor model.  This faster growth
rate was not sufficient to replicate observations of 1987A, however.
In the more recent paper of \citet{Hammer2010}, the authors performed
simulations of a 1987A-type model in two and three dimensions,
starting with a more realistic initial model from \citet{Scheck2007}.
Their initial model was in the first stages of the explosion, and
showed large-scale asymmetry in the heavy element core.  They found,
again, that single perturbations grew around $30\%$ faster in 3D than
in 2D.  The authors proposed that artificial drag forces arising from
2D geometry slow down the growth of the instability in 2D simulations
relative to 3D simulations, where these artificial drag forces are not
present.  This was also described in \citet{Kane2000}, who explain
that the rising bubbles have less effective matter to push out of the
way in 3D than in 2D.

 The \citet{Hammer2010} study also found that in 2D, rising bubbles of
heavy elements became entrained in the dense He shell left behind by
the shock.  In 3D, however, these bubbles were able to burst through
the He shell to reach the H envelope, replicating one of the key
observations of 1987A.

Studies of the Rayleigh-Taylor instability outside the context of
supernova explosions have also found that a single Rayleigh-Taylor
instability grows about $30\%$ faster in 3D than in 2D
\citep{Anuchina2004}.  Previous simulations of the Rayleigh-Taylor
instability in a laboratory context \citep{Miles2005} have found that
the width of the mixed region is very similar between 3D and 2D,
provided that the individual Rayleigh-Taylor fingers interact.
Initially, the Rayleigh-Taylor fingers grow faster in 3D than in 2D.
Once the fingers begin to interact, 3D simulations transition to
turbulence more completely than 2D simulations. 3D simulations can
transfer momentum in two dimensions perpendicular to the blast wave,
while 2D simulations can transfer momentum in only one transverse
direction.  This results in more complete mixing within the mixed
region in 3D than in 2D, reducing the Atwood number in 3D relative to
2D.

For the simplest case of two incompressible fluids under constant
acceleration with an initial perturbation of wavelength $\lambda$ and
amplitude $\eta$, such that $\eta \ll \lambda$, the growth rate of the
instability is given by \citet{Rayleigh1883}:

\begin{equation}
\frac{d\eta}{dt}=\left(\frac{2\pi A g}{\lambda}\right)^{1/2}t
\label{eq:RT}
\end{equation}

where  $g$ is a constant acceleration and $t$ is time.

$A$, the  Atwood number, is defined by

\begin{equation}
A \equiv \frac{(\rho_{2}-\rho_{1})}{(\rho_{2}+\rho_{1})}
\label{eq:At}
\end{equation}
where $\rho_{1}$ and $\rho_{2}$ are the densities of the light and heavy
fluids, respectively.

As the densities of the two fluids become comperable, the Atwood
number becomes smaller and the growth rate decreases.  The growth rate
of the bubbles in 3D is thus reduced, neatly canceling out the reduced
drag these same bubbles experience in 3D as opposed to 2D. The width
of the mixed region remains essentially the same between 2D and 3D
for cases where the Rayleigh-Taylor fingers interact with one another.

It could be supposed that the reason the final state of the supernova
remnant appears the same in our 2D and 3D models is that the
perturbations arising from the grid were somehow larger in 2D than in
3D, and this effect exactly cancelled out the faster growth rates we
expect to see in 3D.  But this is not the case. Perturbations arising
from the grid are larger in 3D than in 2D.  These perturbations arise
because we try to fabricate a spherical object with boxes, as required
by the RZ and Cartesian geometries used in our simulations.  The
maximum amount one would expect to depart from a sphere (or circle) in
space is equivalent to the longest distance across a single cell,
which in 2D is $\sqrt{2}\Delta x$ and in 3D is $\sqrt{3}\Delta x$.  To
make certain, we performed the same set of calculations in 2D at twice
the resolution.  These simulations give the same results as our less
refined simulations, indicating that our simulations are numerically
converged for the problem presented here.

%discussion of why 2D and 3D have same mixing width in our simulations
As in the simulations of \citet{Miles2005}, \citet{Kane2000}, and
\citet{Hammer2010}, we find that the Rayleigh-Taylor instability
initially grows faster in 3D than in 2D.  As the Rayleigh-Taylor
fingers grow to the point where they begin to interact with one
another the simulations begin to transition to a turbulent (3D) or
chaotic (2D) regime. In 3D, momentum is transferred from the fingers
in two dimensions, as opposed to only one dimension in 2D.  The
tranisiton to turbulence in 3D results in more complete mixing within
the mixing region than the 2D simulation, reducing the Atwood number
in 3D relative to 2D.  This has the effect of reducing the growth rate
of the instability in 3D relative to 2D.  The reduced growth rate in
3D relative to 2D at late times compensates for the fact that the
bubbles do indeed experience less drag in 3D than in 2D, and grow
faster at early times.  By the time mixing has ceased, the width of
the mixed region is nearly identical in 2D and 3D in models where the
Rayleigh-Taylor fingers have time to interact with one another.

It can be seen from Eq. \ref{eq:RT} that instabilites seeded by long
wavelength perturbations grow more slowly than those seeded by short
wavelength perturbations.  Obviously, the interior of a supernova
deviates from the ideal case: the acceleration is not constant, the
fluids are compressible, and $\eta$ soon becomes of the same order as
$\lambda$. In order for long wavelength perturbations to dominate
before mixing ceases there must be significant initial power at long
wavelengths.  Because there is less time for the instability to grow
in blue stars, we expect to see more traces of large perturbations at
fairly low wavenumbers in these stars than in red stars.  The
simulations of \citet{Hammer2010} have a significant amount of power
at long wavelengths, so these perturbations grow fast enough to
dominate before mixing is truncated.

%Effect of perturbations on mixing width
The initial perturbations in our simulations are a great deal smaller
than those examined in previous studies.  The spectrum of these
perturbations is also flatter and made up of higher modes.  The
perturbations are more random, though, it should be noted, not
entirely so, and are largest around the 30 and $60^{\circ}$ marks on the
grid.  This can be seen during the initial stages of the development
of the instability, though not at later times, for model s15.  The
enhanced perturbations at these points is likely behind the placement
of the clumps visible in the 3D rendering of model z15 in Figure
\ref{fig:rend}.

%Effect of time and perturbation wavelength on mixing width
Mixing had more time to develop in our two red giant stars, models z15
and s15, than it generally does in the blue giant 1987A-type
progenitor models studied in \citet{Kane2000} and \citet{Hammer2010}.
Mixing froze out after $4x10^{5}$ and $1.7x10^5$ seconds for the red
models s15 and z15, respectively, over an order of magnitude later
than in the simulation of \citet{Hammer2010}, in which mixing froze
out after $9x10^{3}$ seconds.  In our blue u15 model, mixing froze out
after $4x10^4$ seconds, which is still nearly 5 times longer than in
the \citet{Hammer2010} model.  The individual Rayleigh-Taylor
instabilities in all simulations grew for sufficient times that they
interacted, though more interaction took place in the red models.
This was not the case in the single-mode perturbation explored in
\citet{Kane2000}, or the few large perturbation modes in
\citep{Hammer2010}. Smaller wavelengths grow faster than longer
wavelengths, so longer wavelength perturbations are less likely to
become nonlinear. In these models, as well as the single instability
investigated in \citet{Anuchina2004}, the Rayleigh-Taylor fingers did
not interact with one another, ensuring that the relative drag on the
fingers was the only significant difference between 2D and 3D.  This
accounts for the around $30\%$ faster growth rates discovered in these
simulations.  The simplicity of the initial perturbation spectrum can
delay the transition to turbulence, and hence make mixing appear more
efficient in 3D than in 2D \citep{Miles2005}.

For long enough wavelength perturbations, spherical geometry will
ensure that the perturbations diverge, so they never interact and thus
never enter a turbulent regeime.  This will quickly lead to
significant departures from axisymmetry.  There may exist a transition
mode perturbation which would be the longest wavelength mode that
could lead to turbulent interaction.  The instabilities will diverge
without interaction if the wavelength of an instability is greater
than the height at which it would become nonlinear. This tranistion
mode ought to be dependent on presupernova structure: red supergiants
have longer time for the Rayleigh-Taylor instability to grow, and thus
potentially a lower mode than blue giants.

The Rayleigh-Taylor instability may ``forget'' its initial conditions
at late times.  This can happen though an inverse cascade, where
shorter wavelength bubbles merge to form longer wavelength structures.
Under idealized conditions, the flow may become self similar.  This
transition to self-similarity likely happens during the early stages
of Rayleigh-Taylor growth in our simulations, accounting for the
similar end appearance of our simulations despite increasing the
resolution (and decreasing the initial perturbations).  This inverse
cascade can only lead to the instabilities becoming so big. A
necessary condion for self-similarity is that

\begin{equation}
\lambda \ll h \ll L
\end{equation}

where $h$ is the height of the instability and $L$ is the size of the
region in which the instability is growing.  The merger of short
wavelength instabilities to larger wavelength instabilites should be
truncated long before $\lambda$ becomes comperable to $L$, which would
produce large-scale asymmetries.  Large-scale departures from symmetry
must arise from the supernova explosion mechanism itself, as they
cannot arise through an inverse cascade of the Rayleigh-Taylor
instability.

Recent observations suggest that asymmetry in core-collapse type
supernova explosions is common \citep{Wang&Wheeler2008}.  For SNe type
II, this asymmetry is characterized by a dominant polarization angle,
which is most likely due to a directed, jet-like flow.  Polarization
data show that most SNe also exhibit composition-dependent, large
scale departures from axisymmetry.  The degree of departure from
axisymmetry may be less in SNeII-P; \citet{Chugai2005} and
\citet{Chugai2006} are able to match observations of one SNeII-P,
SN2004dj, by using a bipolar distribution of \Ni. This fits with the
emerging picture of collapse supernova mechanisms, which are likely to
be inherently mulAtidimensional and aspherical \citep{Janka2007} The
SASI explosion mechanism \citep{Burrows2006, Scheck2006, Marek2009}
seems to be one of the better candidates for reviving the stalled
shock and successfully exploding a massive star, and it is inherently
asymmetric.  How relevant, then, are spherical explosion models?  This
is the first time the post-explosion hydrodynamics of a wide range of
supernova progenitors has been studied in 3D.  It is reasonable to
model Pop III and Pop II core collapse SNe in 3D first as spherical
explosions as no observations of these primordial supernovae have yet
been made.  This study will contextualize future studies of supernovae
progenitors with asymmetric explosions.

For type II SNe in which departures from asymmetry are small and
dominated by an axisymmetric component, 2D simulations may do as well
as 3D simulations in reproducing the axisymmetric features of the
supernova remnant.  In these stars, Rayleigh-Taylor fingers of all but
the lowest mode order are expected to grow far enough that they begin
to interact with one another.  Simulations in 2D of such stars can be
used to predict nucleosynthesis and fallback for such explosions with
a fair degree of accuracy, as was done in \citet{Joggerst2010}.  2D
calculations may also be used to more accurately predict light curves
of SNe for which departures from axisymmetry do not dominate, which
would use far less computational resources than similar calculations
performed in 3D.

\section{Conclusions}
\lSect{conclusions}

We have performed two and three dimensional simulations of supernova
models with three different metallicities: model z15, with Z = 0
\Zsun, model u15, with Z= $10^{-4}$ \Zsun, and model s15, with
Z=\Zsun. The initial perturbations are seeded by the grid.
Calculations performed at twice the resolution in 2D indicate that our
solutions are numerically converged. We find no significant difference
in the width of the mixed region between our 3D and 2D simulations of
the same supernova.  Previous studies of Rayleigh-Taylor mixing in
1987A-type supernova models found that mixing was more efficient and
individual Rayleigh-Taylor instabilities grew about $30\%$ faster in
3D than in 2D.  \citet{Kane2000} studied single-mode density
perturbations with amplitudes of $10\%$ of mode $l = 10$, and found
that perturbations grew 30-35$\%$ faster in 3D than in 2D.  The more
recent study of \citet{Hammer2010} examined the post-explosion
hydrodynamics of a 1987A-type model with a fully 3D explosion model
from \citet{Scheck2007} with large scale asymmetries at early times.
They found that the instability grew about $30\%$ faster in 3D than in
2D.

While we also find a faster initial growth rate in 3D than in 2D in
our simulations, the growth rate in 3D declines once the simulations
transition to turbulence, and the final width of the mixed region is
the same in 2D and 3D.  Because the Rayleigh-Taylor instability in our
model is seeded by perturbations arising from the grid, our
perturbations have a lower amplitude, and a flatter spectrum shifted
towards higher modes than either the \citet{Kane2000} or
\citet{Hammer2010} studies.  In our progenitor models the
Rayleigh-Taylor instability has more time to develop and grow than in
previous simulations. With a growth time on the order of $10^{5}$
seconds, instabilities in models s15 and z15 grow for more than 10
times longer than the simulations of \citet{Hammer2010}, in which
mixing froze out at around $9 x 10^{3}$ seconds.  In model u15, which
died as a blue supergiant, mixing ceases after $4 x 10^{4}$ seconds,
which is a shorter time than the red giant models in this survey but
is still nearly 5 times longer than the the simulations of
\citet{Hammer2010}.

In our progenitor models and with perturbations seeded by the grid the
RT instability has enough time to grow that the fingers of the
instability begin to interact with one another.  This interaction
transfers momentum and energy transverse to the shock, allowing the 3D
simulations to mix more fully than 2D simulations.  The more efficient
transfer of momentum and energy in 3D renders the 3D simulations fully
turbulent in the mixed region.  This higher degree of mixing in 3D
than in 2D reduces the Atwood number in the 3D simulations relative to
the Atwood number in the 2D simulations.  A lower Atwood number leads
to a reduction in the bubble growth rate in 3D, which compensates for
the lower effective drag in 3D.  The result is that the width of the
mixed region is virtually identical between simulations performed in
2D and those performed in 3D.  3D simulations have transitioned to
full turbulence and are more completely mixed than 2D
simulations. This was found in a laboratory context in the simulations
of \citet{Miles2005}.  For simulations in which interactions between
the Rayleigh-Taylor fingers do not occur, either because only one
finger is modeled \citep{Anuchina2004} or the Rayleigh-Taylor fingers
grow for a short enough time that they never interact
\citep{Kane2000}, the greater effective drag experienced in 2D versus
3D will continue to dominate the simulation, ensuring that the width
of the mixed region will be larger in 3D than in 2D. For simulations
in which the Rayleigh-Taylor instability has enough time to grow that
the fingers begin to interact with one another, the width of the mixed
region will be the same in 2D and 3D simulations.

Instabilities arising from long wavelength perturbations take longer
to grow than those arising from short wavelength perturbations.  Large
amplitude perturbations of long wavelength are necessary for such
instabilities to grow substantially.  These large-scale fingers may
diverge before they can transition to turbulence, and thus may give
rise to departures from axisymmetry.  The large amplitude, long
wavelength perturbations present in the more realistic explosion model
of \citet{Hammer2010} ensure that such fingers don't interact, which
results in the significant differences between their 2D and 3D
simulations.  Rayleigh-Taylor theory predicts that an inverse cascade,
in which shorter wavelength instabilities merge to form longer
wavelength instabilites, should be truncated long before the wavelegth
$\lambda$ becomes comparable to the size of the region in which the
instability is growing.  Large scale departures from symmetry do not
arise through an inverse cascade in our 3D simulations; they must have
their origins in the explosion itself.

It is likely that many, if not most, core-collapse supernovae
explosions are highly asymmetric.  Observations not just of 1987A but
of other supernova, as well the ``kicks'' observed in pulsars,
indicate that asymmetry is a near-universal phenomenon amongst the
explosions of massive stars.  From the theoretical side, it seems
increasingly likely that viable supernova explosions, like the SASI
mechanism, are inherently asymmetric.  In supernovae that die as red
giants the Rayleigh-Taylor instability will have enough time to
develop that individual fingers will interact with one another. If the
asymmetries in such explosions are dominated by axisymmetric
components, 2D models might well be good predictors of the width of
the mixed region and the axisymmetric features of the remnant.  Future
work should investigate asymmetric explosions in a wide variety of
progenitor models.

\acknowledgments

Work at UCSC and LBL was supported in part by the SciDAC Program under
contract DE-FC02-06ER41438.  Work at LANL was carried out under the
auspices of the National Nuclear Security Administration of the
U.S. Department of Energy at Los Alamos National Laboratory under
Contract No. DE-AC52-06NA25396.  Stan Woosley was supported by the NSF
and NASA as well as SciDAC National Science Foundation (AST 0909129)
and the NASA Theory Program (NNX09AK36G).  The simulations were
performed on the open cluster Coyote at Los Alamos National
Laboratory. Additional computing resources were provided on the
Pleiades computer at UCSC under NSF Major Research Instrumentation
award number AST-0521566.  C.C.J. would like to thank Andrew Aspden
for assistance with parallelzing a data reduction routine.  Figure 5
was created using VisIt.

\bibliographystyle{apj.bst}
\bibliography{ms}

\end{document}